**Room Temperature Ferroelectric Ferromagnet in 1D Tetrahedral Chain Network**

*Kyeong Tae Kang, Chang Jae Roh, Jinyoung Lim, Taewon Min, Jun Han Lee, Kyoungjun Lee, Tae Yoon Lee, Seunghun Kang, Daehee Seol, Jiwoong Kim, Hiromichi Ohta, Amit Khare, Sungkyun Park, Yunseok Kim, Seung Chul Chae, Yoon Seok Oh, Jaekwang Lee, Jaejun Yu, Jong Seok Lee, and Woo Seok Choi**

Dr. K. T. Kang, Dr. A. Khare, Prof. W. S. Choi
Department of Physics, Sungkyunkwan University, Suwon 16419, Korea
E-mail: choiws@skku.edu

C. J. Roh, Prof. J. S. Lee
Department of Physics and Photon Science, Gwangju Institute of Science and Technology, Gwangju 61005, Republic of Korea

Dr. J. Lim, Prof. J. Yu
Center for Theoretical Physics, Department of Physics and Astronomy, Seoul National University, Seoul 08826, Republic of Korea

T. Min, J. Kim, Prof. S. Park, Prof. J. Lee
Department of Physics, Pusan National University, Busan 46241, Republic of Korea

J. H. Lee, Prof. Y. S. Oh
Department of Physics, Ulsan National Institute of Science and Technology, Ulsan 44919, Republic of Korea

K. Lee, T. Y. Lee, Prof. S. C. Chae
Department of Physics Education, Seoul National University, Seoul 08826, Republic of Korea

S. Kang, D. Seol, Prof. Y. Kim
School of Advanced Materials Science and Engineering, Sungkyunkwan University, Suwon 16419, Republic of Korea

Prof. H. Ohta
Research Institute for Electronic Science, Hokkaido University, Sapporo 001-0020, Japan

Keywords: $SrFeO_{2.5}$, multiferroicity, ferroelectricity, 1D tetrahedral network, magnetoelectric coupling

Ferroelectricity occurs in crystals with broken spatial inversion symmetry. In conventional perovskite oxides, concerted ionic displacements within a three-dimensional network of transition metal-oxygen polyhedra ($MO_x$) manifest spontaneous polarization. Meanwhile, some two-dimensional networks of $MO_x$ can foster geometric ferroelectricity with magnetism, owing to the distortion of the polyhedra. Because of the fundamentally different mechanism of ferroelectricity in a two-dimensional network, one can further challenge an uncharted mechanism of ferroelectricity in a one-dimensional channel of $MO_x$ and estimate its feasibility. This communication presents ferroelectricity and coupled ferromagnetism in a one-dimensional $FeO_4$ tetrahedral chain network of a brownmillerite $SrFeO_{2.5}$ epitaxial thin film. The result provides a new paradigm for designing low-dimensional $MO_x$ networks, which is expected to benefit the realization of macroscopic ferro-ordering materials including ferroelectric ferromagnets.

Complex transition metal oxides can be analytically viewed as a network of transition metal-oxygen polyhedra ($MO_x$) (**Table 1**). A collective distortion of such networks in a (quasi) two-dimensional (2D) space can lead to inversion symmetry breaking and geometric ferroelectricity through trilinear coupling among distortions and electric polarization.[1-6] In comparison with the conventional displacive-type ferroelectricity in a three-dimensional (3D) space, e.g., relative shifts of the transition metal cation or lone-pair ions, geometric ferroelectricity does not require strong hybridization between the transition metal and oxygen ions.[7-10] This removes the constraint of $d^0$-ness of the transition metal ion, thereby allowing magnetism to coexist with ferroelectricity. In addition, whereas the soft-phonon mode at the Brillouin zone-center manifests displacive ferroelectricity,[11,12] the collective distortion of the 2D network results in zone-boundary soft-phonon modes for geometric ferroelectrics.[3]

What about in one-dimensional (1D) networks? Spatial inversion symmetry breaking can be intuitively envisioned in 1D chain networks as well, especially with an $MO_4$ tetrahedral unit. An $MO_4$ tetrahedron has one of the lowest symmetries among the $MO_x$ polyhedra and can form corner-shared 1D chains within crystals (Table 1). More interestingly, a collective distortion of the 1D network can introduce an unconventional polar state resulting from both the displacement of ions and the rotation of the tetrahedra. A combined polar distortion (CPD) can be considered as a trait of ferroelectricity in 1D channel systems, which can simultaneously introduce ferromagnetism. To realize the CPD within a 1D tetrahedral chain system, we fabricated an epitaxial brownmillerite $SrFeO_{2.5}$ thin film, which shows robust ferroelectricity with inversion symmetry breaking and CPD-induced coupled ferromagnetism.

Brownmillerite $SrFeO_{2.5}$ offers a 1D chain network of $FeO_4$ tetrahedra, in which two stable polarization states can be anticipated.[13-15] Orthorhombic $SrFeO_{2.5}$ possesses alternating layers of $FeO_4$ tetrahedra (blue) and $FeO_6$ octahedra (pink), which stack along the crystallographic $b_o$ (subscript ‘o’ denotes orthorhombic) axis, as schematically illustrated in **Figure 1**a. The $FeO_4$ tetrahedral chains are directed along the $a_o$ axis. The epitaxial strain from the substrate and the thermodynamic parameters during the growth can be employed to determine the orientation of the brownmillerite crystal and $FeO_4$ tetrahedral chains.[16,17] Using pulsed laser epitaxy, we fabricated single-crystalline epitaxial $SrFeO_{2.5}$ thin films on cubic (Nb-doped) $SrTiO_3$ substrates with the following orientation relationships: $SrFeO_{2.5}$ $(\bar{1}01)_o$ // $SrTiO_3$ $(001)_c$ and $SrFeO_{2.5}$ $[010]_o$ // $SrTiO_3$ $[100]_c$ (subscript ‘c’ denotes cubic) (Figure 1b and c). X-ray diffraction (XRD) $\theta$-2$\theta$ scans, $\omega$-scans, reciprocal space maps, and atomic force microscopy validated the expected structural orientation and high crystalline quality (refer Figure S1 and Supporting Information). Within this configuration, the 1D tetrahedral chains are inclined by 45° to the surface normal of the substrate (Figure 1c).

The 1D tetrahedral chain network intrinsically conceives structural instability owing to the CPD provided by the brownmillerite structure. When imagining the rotation of an $MO_x$ polyhedra, the axis of rotation should be defined. For the $MO_6$ octahedra or $MO_5$ triangular bipyramid, the apical oxygen and $M$ ions are located in a line, which naturally serves as the rotation axis. This is not so trivial for the case of $MO_4$ tetrahedra owing to their low structural symmetry. In general, it will be natural to consider the transition metal ion as the center of the rotation. However, within a crystal, e.g., brownmillerite, the alternate stacking of $FeO_6$ octahedral and $FeO_4$ tetrahedral layers anchors the positions of apical oxygen ions, and the line adjoining these ions serves as the rotation axis for each $FeO_4$ tetrahedron (insets of Figure 1d). With this rotation axis defined, CPD occurs naturally in 1D tetrahedral chains with rotational instability (refer Figure S2 and Supporting Information). In addition to the collective polar rotation of the $FeO_4$ tetrahedra, the transition metal ion has to simultaneously undergo a polar displacement with respect to the octahedral layer underneath. This essence of the CPD in the brownmillerite structure presents a stark contrast to the polar distortion in high-dimensional networks, where the polar cation displacement and polar rotation of the polyhedra are incompatible because of the preferential selection of distortive motion.[18]

When the CPDs of each chain exhibit the same handedness (*I*2*bm* space group), polarization will emerge along the 1D tetrahedral chain direction, as shown in the insets of Figure 1d. The polarization value is theoretically estimated to be 7.77 $\mu C/cm^2$ by considering the relative displacement of ions (Figure S2). Figure 1d describes the result of nudged elastic band analysis for the examination of the lowest energy path between the two opposite polarization states, which theoretically supports the CPD-induced ferroelectricity in $SrFeO_{2.5}$. The highly symmetric *Imma* space group is considered as a transition phase. The value of the switching barrier to reverse a tetrahedron is estimated to be ~0.3 eV/tetrahedron. Although this value is larger than the values of the conventional ferroelectrics with tens of meV, it is comparable to geometric ferroelectrics with hundreds of meV.[19]

The CPD implies instabilities corresponding to both polar cation displacement and polar tetrahedral rotation. The two instabilities can be identified by the phonon dispersion relation. Figure 1e and f show the theoretical phonon dispersion curves for the *I*2*bm* (low-temperature phase) and *Imma* (high-temperature phase) structures of $SrFeO_{2.5}$, respectively. Indeed, the *Imma* $SrFeO_{2.5}$ structure exhibits unstable (imaginary) soft modes all across the momentum space, including both the zone center and zone boundary. In particular, multiple dispersion curves can be identified, which might originate from the large structural degree of freedom based on the distortion of a 1D tetrahedral network. The soft-phonon modes at Z- (blue dot) and Γ-points (red dot) are shown in Figure S3b and c, respectively. Both the modes exhibit the combined ionic motions of Fe displacement and tetrahedral rotation expected from previous IR and Raman spectroscopic measurements.[20] Considering the negligible octahedral distortion in the system, the CPD within the tetrahedral chains indeed plays an essential role for the emergent ferroelectricity.[21] We note that the silicate chain in $Bi_2SiO_5$ has been shown to exhibit polar rotation,[22,23] but without the fixed apical oxygen ions as in the brownmillerite structure of $SrFeO_{2.5}$.

The second-harmonic generation (SHG) measurements demonstrate the CPD-induced inversion symmetry breaking in the epitaxial $SrFeO_{2.5}$ thin film (**Figure 2**a-c).[24] Because our thin film is designed to have 1D tetrahedral chains oriented in a direction diagonal to the surface normal (Figure 1), polarization along both the in-plane and out-of-plane directions is anticipated. Clear SHG signals for the $S_{in}$-$S_{out}$ (Figure 2a) and $P_{in}$-$P_{out}$ (Figure 2b) configurations manifest the simultaneous inversion symmetry breaking along the in-plane and out-of-plane directions, respectively. The normal-incident SHG measurement further describes the domain structure with a particular point group symmetry. When defining the normal-incident beam polarization parallel to the $a_c$ and $b_c$ axes as $X$ and $Y$, respectively, the $\varphi$-dependent SHG intensity of $X_{in}$-$X_{out}$ and $X_{in}$-$Y_{out}$ exhibits a four-fold symmetry, as shown in Figure 2c. The particular shape of the $\varphi$-dependent SHG result can be well fitted based on the point group symmetry $mm2$ of the expected $I2bm$ structure, for both the $X_{in}$-$X_{out}$ and $X_{in}$-$Y_{out}$ configurations. The fitting suggests an area ratio of 4.6:5.4 between the domains of $b_o$ // $a_c$ and $b_o$ // $b_c$, which stems from the structural degeneracy offered by the cubic substrate. Indeed, the macroscopic (micrometer)-scale structural domains with a phase difference of 90° are clearly visible on the surface of the $SrFeO_{2.5}$ thin film in the polarized optical microscope images (refer Figure S4g and Supporting Information). Although the domain structure might reduce the net magnitude of the in-plane polarization, the out-of-plane polarization is expected to remain robust.

The CPD-induced spatial inversion symmetry breaking leads to switchable ferroelectricity. Figure 2d shows the hysteretic polarization versus electric field curve of the $SrFeO_{2.5}$ epitaxial thin film, indicating robust ferroelectricity at room temperature (black line). The loop is obtained by the junction switching current (red line), which shows a peak near the coercive field (details in Experimental Section). Despite the leaky nature of the switching current, a remnant polarization of 3 $\mu C/cm^2$ with a large coercive field of 1.5 MV/cm is obtained, which is consistent with the theoretical estimation. In addition, piezoresponse force microscopy (PFM) was employed to minimize the effect of the leakage current. Figure 2e shows the apparent hysteresis loop of the piezoresponse with a coercive field of 1.25 MV/cm, which is consistent with the switching current measurement. The negative correlation between the loop size and temperature in the temperature-dependent PFM measurement eliminates the possible contribution from possible ionic conduction to the loop.[25,26] The piezoresponse amplitude and phase (Figure 2f and g, respectively) and the bipolar box pattern images for the amplitude and phase (Figure 2h and i, respectively) further yield clear ferroelectric characteristics.

The CPD in a 1D tetrahedral network promotes ferromagnetic ordering which is coupled to the polar order. In particular, the alternate displacive motion of Fe and O in the typically antiferromagnetic brownmillerite $SrFeO_{2.5}$ leads to the Dzyaloshinskii-Moriya (DM) exchange interaction, as illustrated in **Figure 3**a-c.[1,27] A finite Dzyaloshinskii vector ($\mathbf{D}_{ij}$), resultant from the CPD-induced ferroelectricity, stabilizes the ferromagnetic ground state (refer Supporting Information). Figure 3d demonstrates the hysteretic magnetization versus magnetic field at 300 K. A clear ferromagnetic hysteresis with a small coercive field of ±130 Oe (dashed vertical lines) and remnant magnetization of ~0.01 $\mu_B$/Fe is observed, indicating canted ferromagnetism. The magnetoelectric coupling between the polar and magnetic orders via CPD is quantitatively evaluated by AC magnetoelectric coupling measurements. Figure 3e shows the change in the AC magnetoelectric coupling constant $\Delta\alpha$ as a function of the DC magnetic field $H$ at 300 K. The AC magnetoelectric coupling constant $\alpha$ is defined as $\alpha \equiv \delta P_{ac}/\delta H_{ac} = \delta Q_{ac}/(\delta H_{ac} \cdot A)$, where $\delta P_{ac}$ and $\delta Q_{ac}$ are the amplitude of modulated electric polarization and charge by the applied AC magnetic field $\delta H_{ac}$, respectively, and $A$ is the area of the electrode. Eddy current (induced by the Faraday's law) contribution to $\alpha$ was ruled out by considering a 90° phase difference relative to the $H_{ac}$ (Figure S5b). $\Delta\alpha$ was obtained by subtracting constant $\alpha$ values from $\alpha(H)$ (Figure S5). While $\alpha$ itself is already a straightforward indicator of the magnetoelectric coupling, the underestimated $\Delta\alpha$ exhibits a clear magnetoelectric hysteresis with the peaks feature corresponding to the ferromagnetic coercive fields, indicating the synchronized reversal of the polar and magnetic domains driven by the external magnetic field. Note that a relatively small $\Delta\alpha$ is ascribed to the compensated magnetoelectric effect, which results from multiple magnetoelectric domains and/or the thermal instability of structural distortion. This result manifests the realization of magnetoelectrically coupled ferroelectric ferromagnets at room temperature.[28] Thus, 1D tetrahedral networks provide an unprecedented platform for the engineering of low-dimensional nanoscale networks, thereby leading to coupled ferro-ordering.

**Experimental Section**

*Thin film growth and structural characterization*: Epitaxial $SrFeO_{2.5}$ thin films were grown on atomically flat (001)-oriented single-crystalline $SrTiO_3$ substrates using pulsed laser epitaxy at 700 °C under 1 mTorr of oxygen partial pressure. For the electrical measurements, metallic Nb-doped (0.5 wt%) $SrTiO_3$ substrate was employed as the bottom electrode. An excimer (KrF) laser of 248 nm wavelength (IPEX 864, Lightmachinery, Nepean, Canada) with an energy fluence of 1.3 J $cm^{-2}$ and a repetition rate of 4 Hz was used. The atomic structures and crystallinity of the epitaxial thin films were characterized using high-resolution XRD. The thickness of the thin films was determined as 36 nm, using X-ray reflectivity.

*Density functional theory calculation*: Density functional theory calculations were performed using the Vienna *Ab initio* simulation package (VASP).[29] We employed the generalized gradient approximation-Perdew Burke Ernzerhof (GGA-PBE) exchange-correlation functionals[30] and the projected augmented wave pseudopotentials.[31] We included the on-site Coulomb repulsion for the Fe *d* electrons using the GGA+$U$ method with the effective Hubbard interaction parameter of $U$ = 3 eV in the Dudarev implementation.[32] The energy barrier to invert the polarization to the opposite direction was calculated using the nudged elastic band (NEB) method[33] as implemented in VASP, and the macroscopic polarization was obtained by the Berry phase method.[34] Phonon calculations were employed using the PHONOPY code.[35,36] To activate non-analytical term correction for improving the precision, the Born effective charges and dielectric constants were obtained using the density functional perturbation theory.[37] The NEB and Berry phase calculations were performed with the energy cutoff for a plane wave basis set of 600 eV, and we further used a brownmillerite supercell including 36 atoms corresponding to $\sqrt{2} \times \sqrt{2} \times 4$ perovskite cells with $6 \times 6 \times 2$ Monkhorst-Pack $k$-point grid. The lattice structures were relaxed until the forces converged within 0.01 eV/Å, while the energy criterion was $10^{-6}$ eV. Meanwhile, for the phonon calculations, both the *I*2*bm* and *Imma* phases were optimized with an energy cutoff of 500 eV and Γ-centered 4 × 4 × 2 *k*-point grid meshes. The lattice structures were fully relaxed until the forces were less than $10^{-6}$ eV/Å and converged in energy to $10^{-9}$ eV/unit cell.

*SHG measurements*: SHG is an efficient method to investigate the broken inversion symmetry nature of a material non-destructively. It is feasible because the non-zero components of the second-order susceptibility ($\chi^{(2)}$) tensor reflect the absence of spatial inversion symmetry in a medium.

We monitored the second-harmonic response at room temperature using a Ti-sapphire oscillator (Vitara-T) with a central wavelength of 800 nm. A femtosecond laser with a pulse width of 30 fs and a repetition rate of 80 MHz was used. We focused on the fundamental wave with a beam size of approximately 100 $\mu m^2$ using a plano-convex lens with a focal length of 100 mm. Depending on the purpose, both the grazing incidence with an incident angle of 45° and the normal incidence were exploited. The full-angle variations of the polarization of the fundamental wave and second-harmonic wave were controlled using a half-wave plate and polarizer, respectively, by which the *P*- or *S*-wave was well defined. The second-harmonic intensity was detected using a photomultiplier tube.

*Ferroelectric polarization measurement and PFM*: Polarization measurements were performed using a TF analyzer 3000 (aixACCT systems Co., Germany) at room temperature. Electric pulses of 2 kHz were applied to switch and measure the polarization of the thin film. The junction conductance was measured for a Cr, Au//$SrFeO_{2.5}$/Nb (0.5 wt %):$SrTiO_3$ device. The Cr/Au pad and Nb:$SrTiO_3$ substrate served as the top and bottom electrodes, respectively, as depicted in the inset of Figure 2d. From the junction conductance, the polarization can be obtained by

$$I(t) = A \times \frac{dP(t)}{dt} + C \times \frac{dV(t)}{dt} + I_c, \quad (1)$$

where $I(t)$ and $V(t)$ are the junction current and voltage, respectively, and $I_C$ is the conductance current.[38] We exploited a high frequency of 2 kHz, which is required to measure a leaky thin film sample owing to the non-zero $3d^5$ electrons with a small band gap of ~1.5 eV. To overcome the large conductance current during the direct polarization measurement, we also used a tip-based approach to firmly determine the ferroelectricity. The piezoresponse measurements were performed using a commercial atomic force microscope (AFM) system (Park Systems, NX-10) equipped with a function generator and data acquisition systems (National Instruments, NI-PXIe 5122/5412) (inset of Figure 2e). The piezoresponse was measured by applying a DC voltage ranging from −8 to 8 $V_{DC}$ with a band excitation waveform of 2 $V_{AC}$ between 320 and 400 kHz to a conductive Pt/Cr-coated AFM tip (Budget Sensors, Multi75E-G) with a spring constant of 3 N/m. A commercial AFM heating stage (Park Systems, Temperature Control Stage 1) was used to increase the temperature of the samples during the measurements. A switchable polarization of the $SrFeO_{2.5}$ epitaxial thin film was observed with a clear hysteresis behavior of the $d_{33}$ signal from the PFM measurements after applying −8 V to the tip at room temperature, as shown in Figure 2e. Figure 2f and g show the piezoresponse amplitude-voltage butterfly curve and hysteretic phase loop of the thin films, respectively. Furthermore, the surface of the thin film was scanned with an electrically biased tip of ±6 V, in a 10 × 10 μ$m^2$ area for a box-patterned poling. The negative (centre of box) and positive (box-border) poling regions are clearly distinguished in the amplitude (Figure 2h) and phase (Figure 2i) maps.[39,40]

*Characterization of magnetic properties in thin films*: Magnetization measurements for the $SrFeO_{2.5}$ thin film were performed using a commercial SQUID magnetometer (MPMS-3, Quantum Design Inc.) by applying a maximum field of 0.5 T at 300 K. The magnetic moment was obtained by vibrating the samples at 12.8 Hz (VSM mode) using a scan length of 5 mm and an average time of 2 s. Measurement for the magnetic property of the thin film is schematically shown in the inset of Figure 3d.

*Quantitative determination of magnetoelectric coupling*: In magnetoelectric materials, oscillation of externally applied magnetic field ($\delta H_{ac}$) produces modulation of electric polarization ($\delta P_{ac}$). The magnetoelectric coupling constant $\alpha$ is defined by $\alpha = \delta P_{ac}/\delta H_{ac}$. The unit of $\alpha$ is s/m, which comes from the unit of $P_{ac}$ in C/m$^2$ and of $H_{ac}$ in A/m = C/(m·s). To determine $\alpha$ quantitatively, we developed a homemade magnetoelectric susceptometer. The physical property measurement system (Quantum Design) was employed to apply a DC magnetic field $H$. The AC magnetic field was applied by a pair of counterwound solenoid coils, which compensates for the stray magnetic field outside the solenoid coils and minimizes the magnetic field induced electrical noise. A three-turned coil, which was calibrated to the AC magnetic field as a function of applied AC current and frequency, was employed to determine $\delta H_{ac}$. The gold electrode was used as the top electrode, while the Nb-doped $SrTiO_3$ substrate was used as the bottom electrode. The sample was positioned at the center of the solenoids. By oscillating $\delta H_{ac}$ at a frequency of 1 kHz while applying $H$, electric polarization within the epitaxial film modulates with the same frequency of 1 kHz and an amount of $\delta P_{ac} = \delta Q_{ac}/A$, where $\delta Q_{ac}$ and $A$ are the amount of charge modulation and the area of the top electrode, respectively. $\delta Q_{ac}$ was amplified and converted to the calibrated voltage signal $\delta V_{ac}$ by a homemade charge amplifier and lock-in amplifier (Stanford Research SR830). Thus, $\alpha$ was estimated by $\alpha \equiv \delta P_{ac}/\delta H_{ac} = \delta Q_{ac}/(\delta H_{ac} \cdot A)$. For the transverse $\perp$ and longitudinal // configurations, both AC and DC magnetic fields were applied along the in-plane and out-of-plane of the thin film, respectively. To rule out possible extrinsic contributions, such as thermally excited carriers and eddy current on the electrode for the measured values of $\alpha$, we present $\Delta\alpha$, where constant values of $\alpha_{\perp,0}$ and $\alpha_{//,0}$ were subtracted. It manifests the explicit hysteretic behavior as illustrated in Figure 3e, despite the modest values of as/m = $10^{-18}$ s/m.

**Supporting Information**
Supporting Information is available from the Wiley Online Library or from the author.

**Acknowledgements**

We thank G. Catalan, H. Taniguchi, C. H. Kim, J. M. Rondinelli, and J. Young for their insightful discussion. This work was supported by the Basic Science Research Programs through the National Research Foundation of Korea (NRF) (NRF-2019R1A2B5B02004546, 2017R1A4A1015323, and 2018M3D1A1089406).

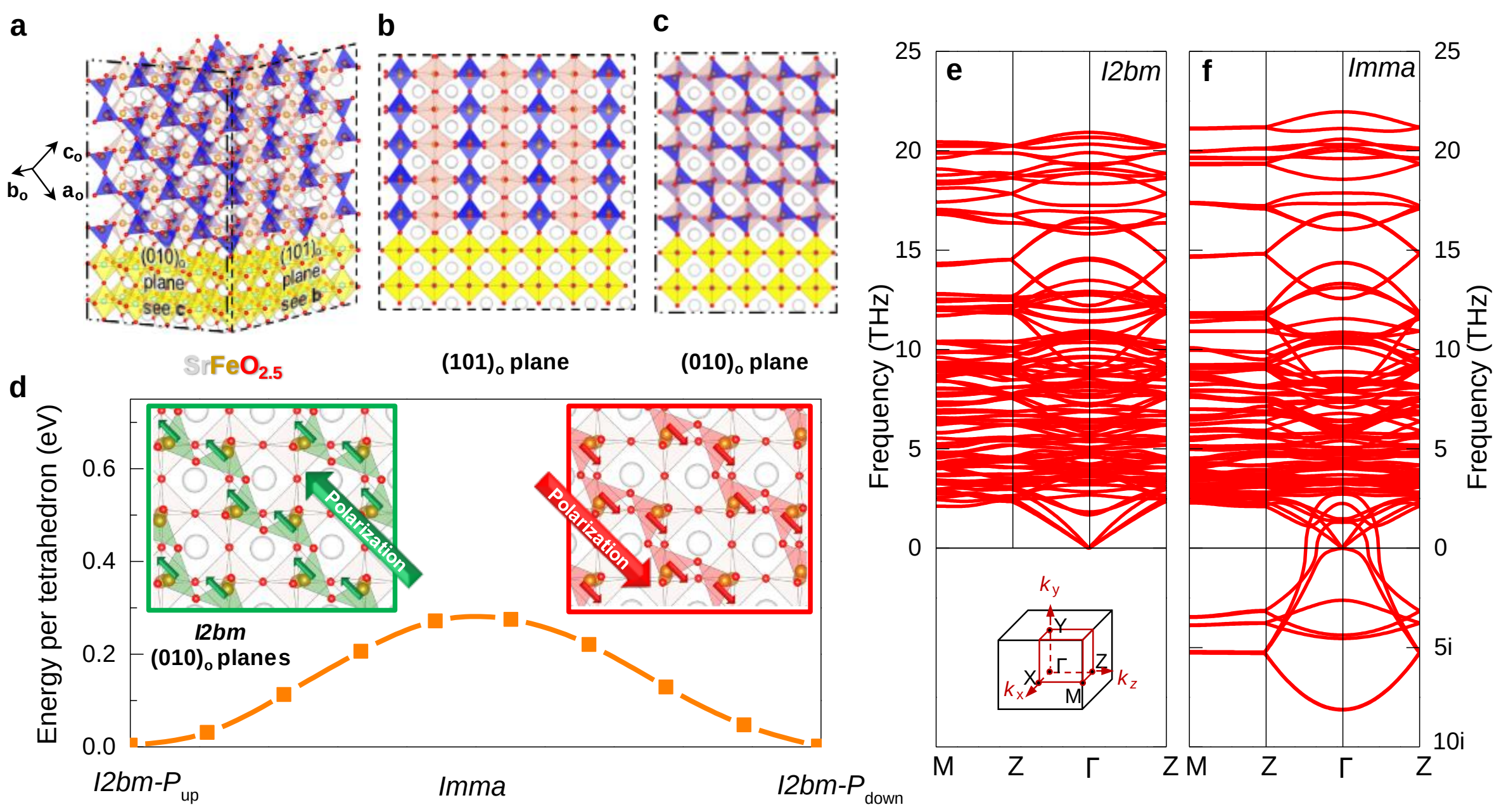

**Figure 1.** Brownmillerite $SrFeO_{2.5}$ with CPD in 1D tetrahedral chains. (a) Schematic diagram of orthorhombic $SrFeO_{2.5}$ epitaxial thin film on a $SrTiO_3$ substrate with alternating $FeO_4$ tetrahedral (blue) and $FeO_6$ octahedral (pink) layers. Cross-sectional schematic views for the (b) $(101)_o$ and (c) $(010)_o$ planes. (d) Minimized energetic path to reverse the 1D tetrahedral chain. The *I2bm* structure of $SrFeO_{2.5}$ in the left and right insets represents a polarized state, and the non-polar symmetric *Imma* structure serves as a transition state. Phonon dispersion curves for (e), ferroelectric *I2bm*, and (f), unstable *Imma* $SrFeO_{2.5}$ structures.

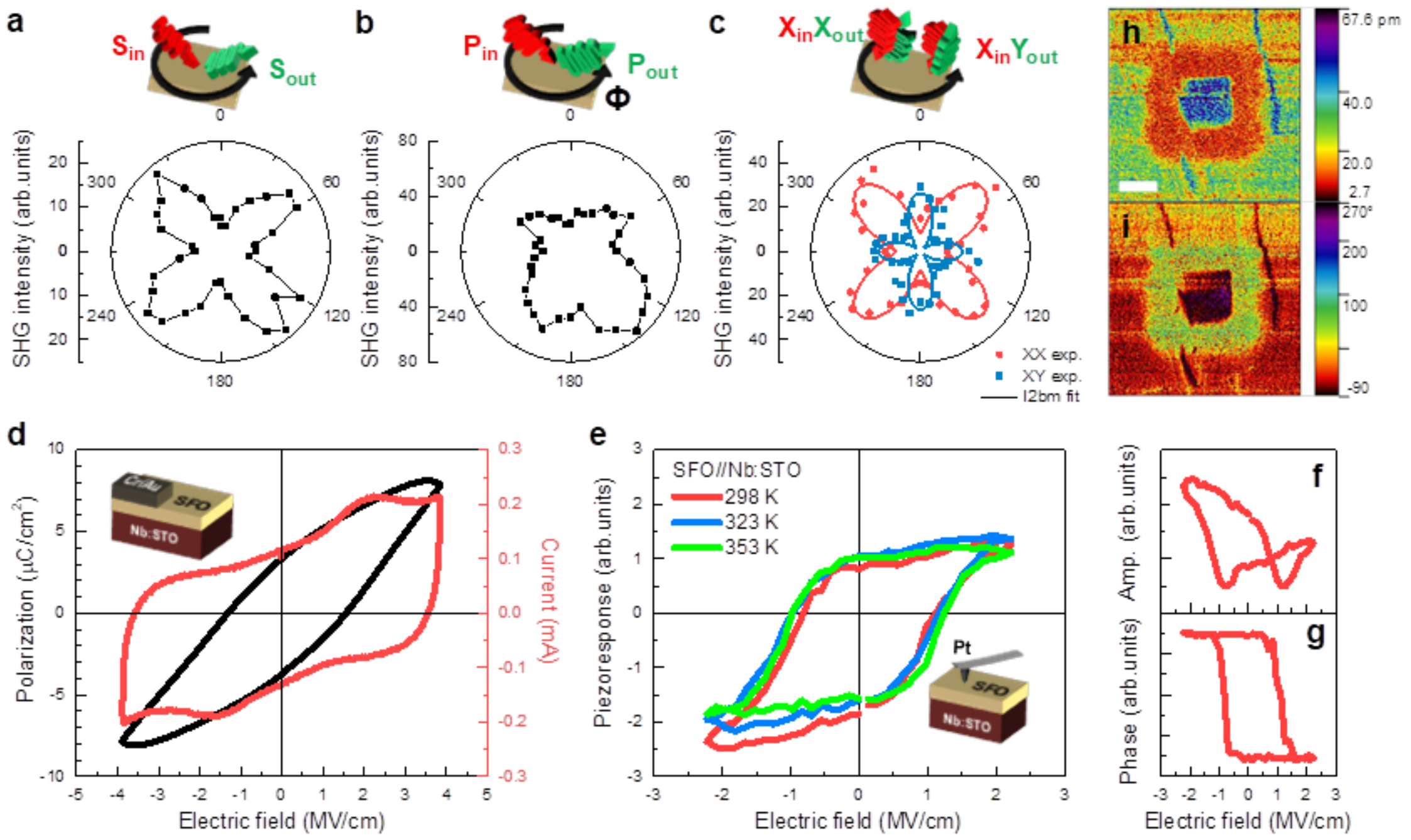

**Figure 2.** Ferroelectricity in the brownmillerite epitaxial thin film. (a) $S_{in}$-$S_{out}$ (in-plane) and (b) $P_{in}$-$P_{out}$ (out-of-plane symmetry breaking) configuration of the SHG signal. (c) Normal-incident SHG signal (dots) and its fit (solid lines) based on the *I2bm* $SrFeO_{2.5}$, where the $a_c$-parallel and $b_c$-parallel polarization geometries are denoted as X and Y, respectively. (d) Polarization (black) and junction current (red) as a function of electric field for the epitaxial $SrFeO_{2.5}$ thin film at room temperature. (e) Hysteresis loop obtained by PFM measurement at 298, 323, and 353 K. The insets show the schematic experimental configurations. Piezoresponse (f) amplitude and (g) phase as a function of electric field. Box-patterned PFM images of (h) amplitude and (i) phase.

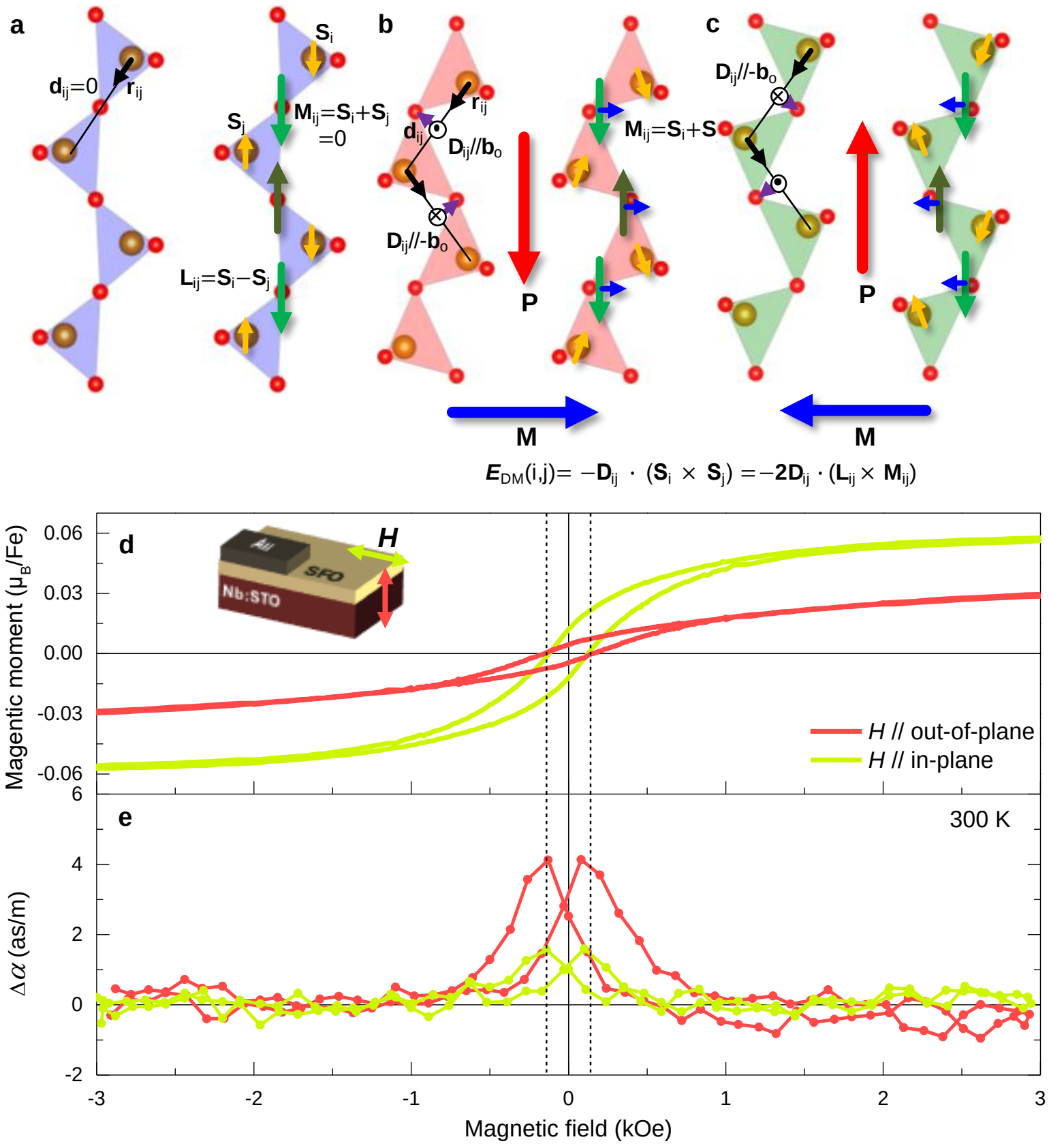

**Figure 3.** Room temperature ferromagnetism coupled with polarization. $\mathbf{r}_{ij}$ (black line) is the vector pointing towards a magnetic ion from a neighboring magnetic ion. $\mathbf{d}_{ij}$ (purple arrow) is the shift vector, which can be defined as the shift of oxygen ion from the line connecting magnetic ions. $\mathbf{L}_{ij}$ (= $\mathbf{S}_i - \mathbf{S}_j$, green arrows) is an antiferromagnetic vector defined as the difference between the spin vectors ($\mathbf{S}_i$ and $\mathbf{S}_j$, yellow arrows) of neighboring sites. $\mathbf{D}_{ij}$ is the Dzyaloshinskii vector pointing either towards the page or outwards from the page. $\mathbf{M}_{ij}$ (= $\mathbf{S}_i$ + $\mathbf{S}_j$, blue arrows) is the local magnetic moment. (a) When CPD is not present, $\mathbf{d}_{ij}$ = 0; hence, $\mathbf{D}_{ij}$ and $\mathbf{M}_{ij}$ are all zero for the antiferromagnetically ordered spins parallel to the chain direction. (b and c) When CPD occurs, the distortion within the tetrahedral chains gives rise to a finite $\mathbf{d}_{ij}$; hence, a collinear canting of the spins can be expected via the DM interaction. The resultant magnetization **M** is closely coupled to the electric polarization **P**. (d and e) Ferromagnetism of epitaxial $SrFeO_{2.5}$ thin film magnetoelectrically coupled with electric polarization at room temperature. For both out-of-plane (red line) and in-plane (green line) magnetic fields, a magnetic hysteresis $M(H)$ and corresponding variation of magnetoelectric coupling constant $\Delta\alpha(H)$ are observed as a function of the applied DC magnetic field $H$. For $\Delta\alpha(H)$, constant values of $\alpha_{//,0}$ and $\alpha_{\perp,0}$ were subtracted from $\alpha(H)$, as shown in Figure S5. The vertical dashed lines represent the coercive field.

| Network Dimension | Unit polyhedron | Ferroelectric mechanism | Representative materials |
|---|---|---|---|
| **3D network** | Octahedron[a] | Hybridization or lone pair | PZT, $BaTiO_3$, $BiFeO_3$, $BiMnO_3$ |
| | | Charge ordering | $Fe_3O_4$ |
| | | Magnetic ordering | orthorhombic-*R*$MnO_3$ (*R* = Gd, Tb, Dy) |
| **2D network** | Octahedron[b] | Geometric distortion | Ruddlesden-Popper phases, $PbTiO_3$/$SrTiO_3$ superlattice, $Ca_3Ti_2O_7$, $K_2SeO_4$, $Cs_2CdI_4$ |
| | Triangular bipyramid[c] | Geometric distortion | Hexagonal-*R*$MnO_3$ (R = Ho-Lu, Y), $LuFeO_3$ |
| | | Charge ordering | $LuFe_2O_4$ |
| **1D channel** | Tetrahedron[d] | Combined polar distortion | **$SrFeO_{2.5}$** |

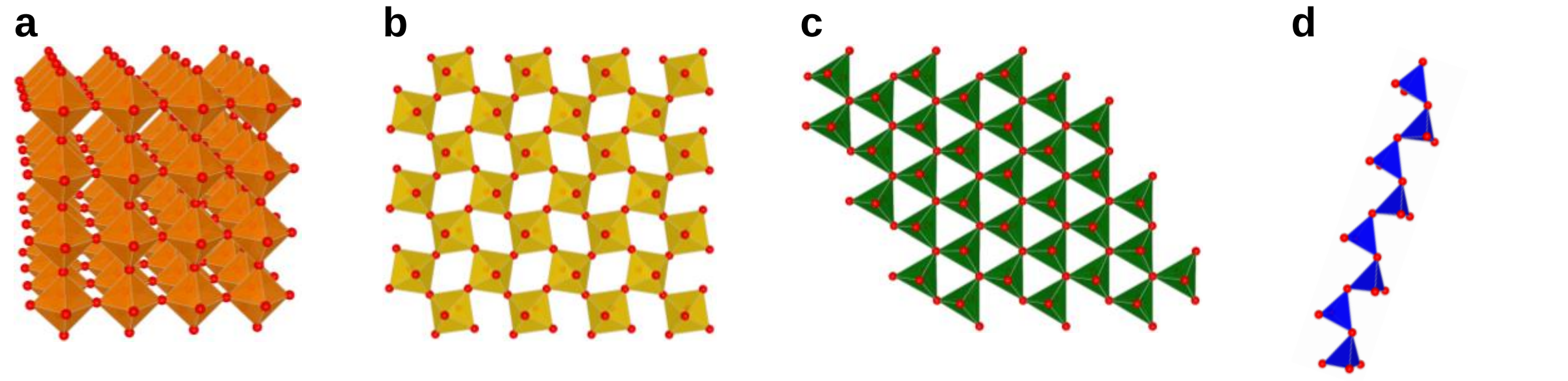

**Table 1.** Classification of transition metal oxide ferroelectrics based on the dimension of the polyhedra networks.

# Supporting Information

**Room Temperature Ferroelectric Ferromagnet in 1D Tetrahedral Chain Network**

*Kyeong Tae Kang, Chang Jae Roh, Jinyoung Lim, Taewon Min, Jun Han Lee, Kyoungjun Lee, Tae Yoon Lee, Seunghun Kang, Daehee Seol, Jiwoong Kim, Hiromichi Ohta, Amit Khare, Sungkyun Park, Yunseok Kim, Seung Chul Chae, Yoon Seok Oh, Jaekwang Lee, Jaejun Yu, Jong Seok Lee, and Woo Seok Choi**

## Orientation of the $SrFeO_{2.5}$ epitaxial thin film

To realize and observe 1D combined polar distortion (CPD) in a brownmillerite $SrFeO_{2.5}$ thin film, the characterization of the crystallographic orientation of the thin film should precede. A first-principles calculation study and an independent experimental observation reveal that the orthorhombic unit cell of brownmillerite $SrFeO_{2.5}$ can be stabilized into two states with different orientations on the $SrTiO_3$ substrate. That is, $SrFeO_{2.5}$ $[010]_o$ // $SrTiO_3$ $[100]_c$ and $SrFeO_{2.5}$ $[010]_o$ // $SrTiO_3$ $[001]_c$ are nearly degenerate. Here, the subscript 'o' and 'c' denote orthorhombic and cubic, respectively. To observe ferroelectricity, we selected the former condition (left inset of Figure S1a and Figure 1a-c), in which the 1D tetrahedral chain network is 45° tilted from the surface normal of the thin film. The X-ray diffraction (XRD) $\theta$-$2\theta$ scan confirmed the case of $SrFeO_{2.5}$ $[010]_o$ // $SrTiO_3$ $[100]_c$, as shown in Figure S1a. The XRD reciprocal space map (Figure S1b) shows the three expected diffraction peaks of the $SrFeO_{2.5}$ thin film, again corresponding to the crystallographic orientation of $SrFeO_{2.5}$ $[010]_o$ // $SrTiO_3$ $[100]_c$. The right inset of Figure S1a shows the atomic force microscopy (AFM) image of the atomically flat surface of the $SrFeO_{2.5}$ epitaxial thin film.

## CPD within the $SrFeO_{2.5}$ tetrahedral chains

The polarization in the brownmillerite $SrFeO_{2.5}$ stems from the spatial inversion symmetry breaking, which is attributed to the mono-handedness distortion of 1D tetrahedral chains. Comparing the ionic shift between the highly symmetric *Imma* and polar *I*2*bm* states provides the essential feature that drives the polar nature in $SrFeO_{2.5}$. First, the Sr ions and $FeO_6$ cages in the octahedral layers exhibit negligible displacement, indicating their incompetence to generate a finite electric polarization. On the other hand, the oxygen and iron ions of the $FeO_4$ tetrahedra layer show significant displacements with respect to the octahedral layers, as depicted in Figure S2. The rotation of the asymmetric tetrahedra, whose rotation axis of apical oxygen ions is not in line with Fe, simultaneously introduces both the rotation of $FeO_4$ tetrahedra and the ionic displacement of Fe. Remarkably, the chain-parallel components of the polarization survive for each chain, whereas the chain-perpendicular components are averaged out.

**Domain structure characterization via second-harmonic generation**

To characterize the domain structure, the configuration with normal incidence second-harmonic generation (SHG) measurement was exploited. Although the four-fold shapes in the $S_{in}$-$S_{out}$ and $P_{in}$-$P_{out}$ SHG investigations already confirmed the existence of broken inversion symmetry domain within the thin film, normal incidence SHG measurements further provided precise information of the domain. The measurements were performed with $a_c$-polarized (denoted by $X$) and $b_c$-polarized (denoted by $Y$) beams defined at $\varphi = 0$. Figure S4 illustrates the simulated SHG results for different orientations of the domains using the $X_{in}$-$X_{out}$ and $X_{in}$-$Y_{out}$ configurations along with the experimental results. Figure S4a and b depict the simulations for $X_{in}$-$X_{out}$ and $X_{in}$-$Y_{out}$, respectively, of a $SrFeO_{2.5}$ $[010]_o$ // $SrTiO_3$ $[100]_c$ single domain. Figure S4c and d depict the same simulations, respectively, of a domain mixture with half of $SrFeO_{2.5}$ $[010]_o$ // $SrTiO_3$ $[100]_c$ and half of $SrFeO_{2.5}$ $[010]_o$ // $SrTiO_3$ $[010]_c$. The combination of these simulations provides a reasonable model, which fits well with the experimental results (Figure S4e and f), indicating that the $SrFeO_{2.5}$ thin film possesses both oriented domains owing to the energetic degeneracy. The result is consistent with polarized optical microscope images (Figure S4g) in which the reversed color contrast for orthogonal incident light polarization yields micrometer-sized structural domains with a 90° phase difference.

**Canted ferromagnetism**

Although $SrFeO_{2.5}$ is originally an antiferromagnetic insulator, a ferromagnetic ordering can be introduced via the Dzyaloshinskii-Moriya exchange mechanism. In particular, the alternate shift of oxygen ions from the line connecting the transition metals within the 1D chain can lead to a collinear canting of spins, thereby leading to a non-zero net magnetization. Without the CPD, brownmillerite $SrFeO_{2.5}$ carries the antiferromagnetic spin vectors that are parallel to the tetrahedral chain, shown as yellow arrows in Figure 3a. Therein, the antiferromagnetic vectors $\mathbf{L}_{ij}$ (green arrows), defined as the difference in the neighboring spin vectors $\mathbf{S}_i$ and $\mathbf{S}_j$, are along the tetrahedral chain. Meanwhile, the local magnetic moments $\mathbf{M}_{ij} = \mathbf{S}_i + \mathbf{S}_j$ are zero. Moreover, the shift in the oxygen ions from the line connecting the neighboring magnetic ions ($\mathbf{r}_{ij}$, black line) does not occur ($\mathbf{d}_{ij} = 0$) in the case without CPD. When CPD is turned on within the tetrahedral chains (Figure 3b and c), the oxygen ions shift; thus, leading to a finite shift vector ($\mathbf{d}_{ij}$, purple arrows). Because the Dzyaloshinskii vector $\mathbf{D}_{ij}$ is proportional to $\mathbf{d}_{ij} \times \mathbf{r}_{ij}$, all the antiferromagnetic spin states experience canting through the interaction $H_{DM} = 2\mathbf{D}_{ij} \cdot (\mathbf{L}_{ij} \times \mathbf{M}_{ij})$. Here, $\mathbf{M}_{ij}$ is the local magnetic moment (blue arrows). Owing to the close correlation between the CPD handedness and the direction of polar ordering, the resultant magnetization $\mathbf{M}$ should be coupled with polarization $\mathbf{P}$.

## SUPPORTING FIGURE LEGENDS

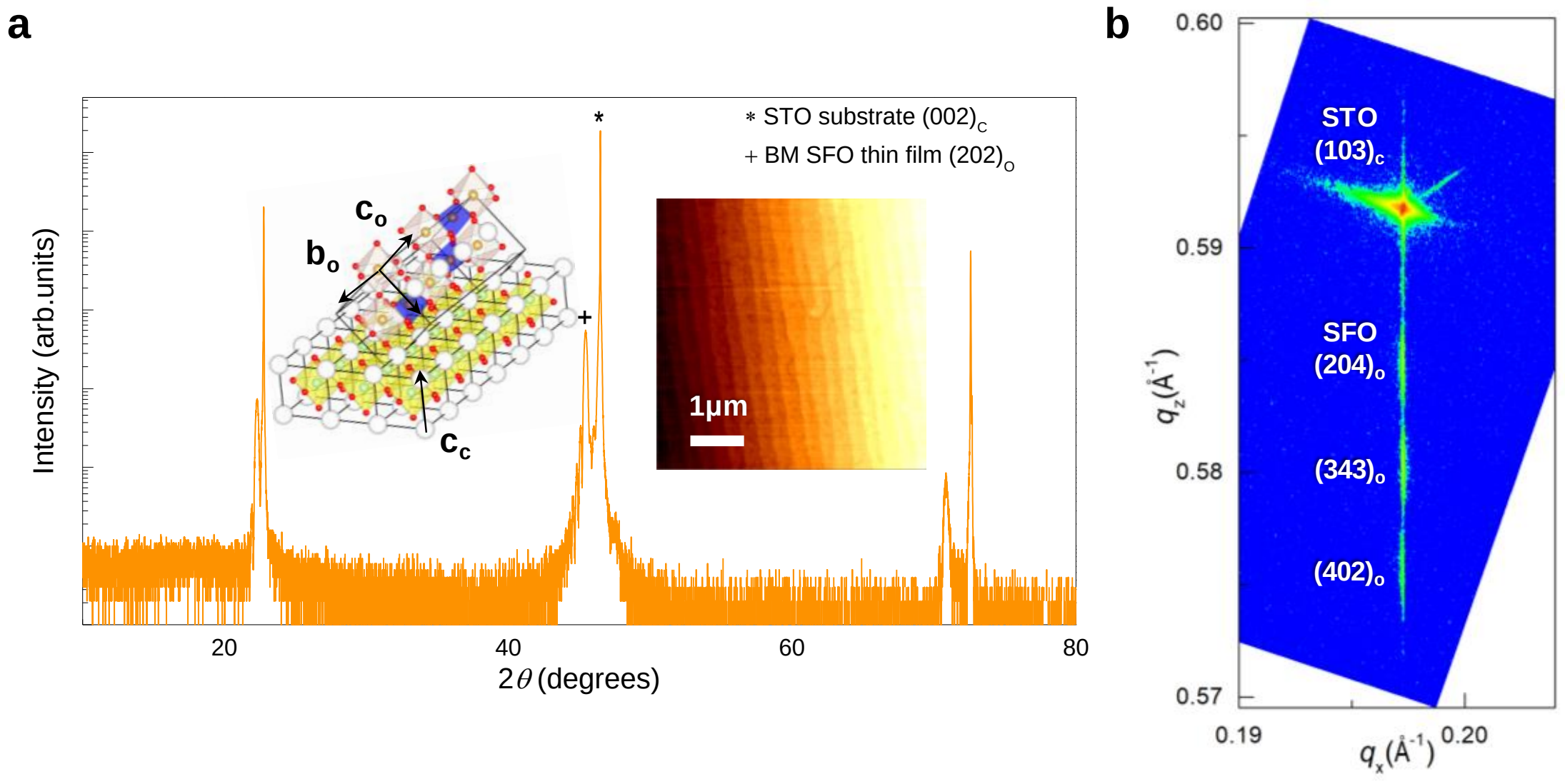

**Figure S1.** $SrFeO_{2.5}$ epitaxial thin film characterization. (a) XRD $\theta$-$2\theta$ scan result of the $SrFeO_{2.5}$ epitaxial thin film, indicating a highly crystalline brownmillerite structure with the orientational relationship of $SrFeO_{2.5}$ $[010]_o$ // $SrTiO_3$ $[100]_c$. The left inset describes the schematic diagram of an orthorhombic $SrFeO_{2.5}$ unit cell deposited on a cubic $SrTiO_3$ substrate. The right inset shows the AFM topography image of the epitaxial thin film surface with preserved one-unit-cell step and terrace structure of the substrate. (b) The XRD reciprocal space map of $SrFeO_{2.5}$ epitaxial thin film on the $SrTiO_3$ substrate around the (103) Bragg reflection of $SrTiO_3$. The three peaks of the thin film explicitly confirm the orientation relationship of $SrFeO_{2.5}$ $[010]_o$ // $SrTiO_3$ $[100]_c$.

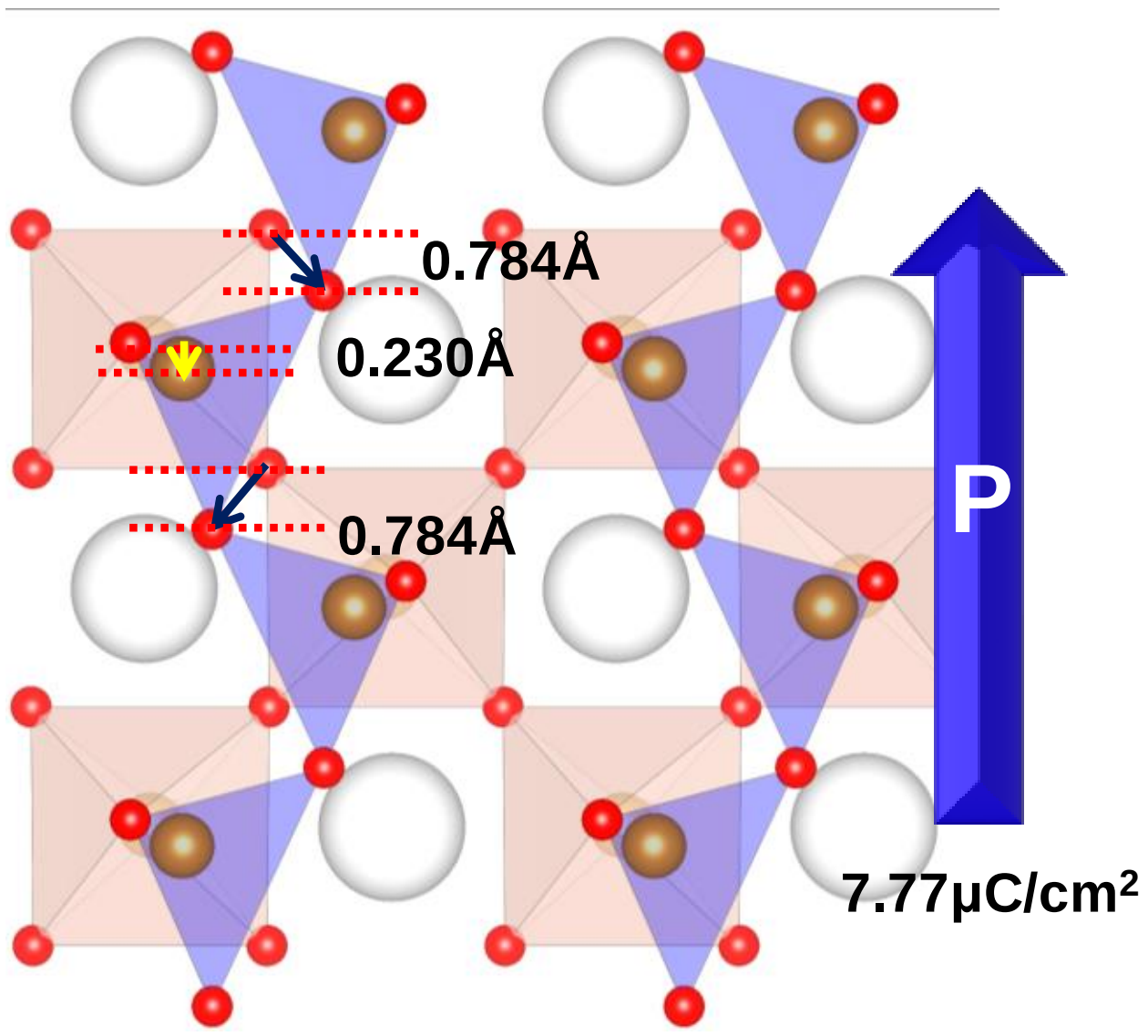

**Figure S2.** Ionic displacements in $SrFeO_{2.5}$ via CPD. The structural modification from a non-polar *Imma* structure to a polar *I*2*bm* structure is captured by the collective distortion of tetrahedral chains. While the ions within the octahedral layer do not move significantly, the ions in the tetrahedral layer show significant displacement corresponding to the spontaneous polarization of 7.77 $\mu C/cm^2$ along the 1D chain direction.

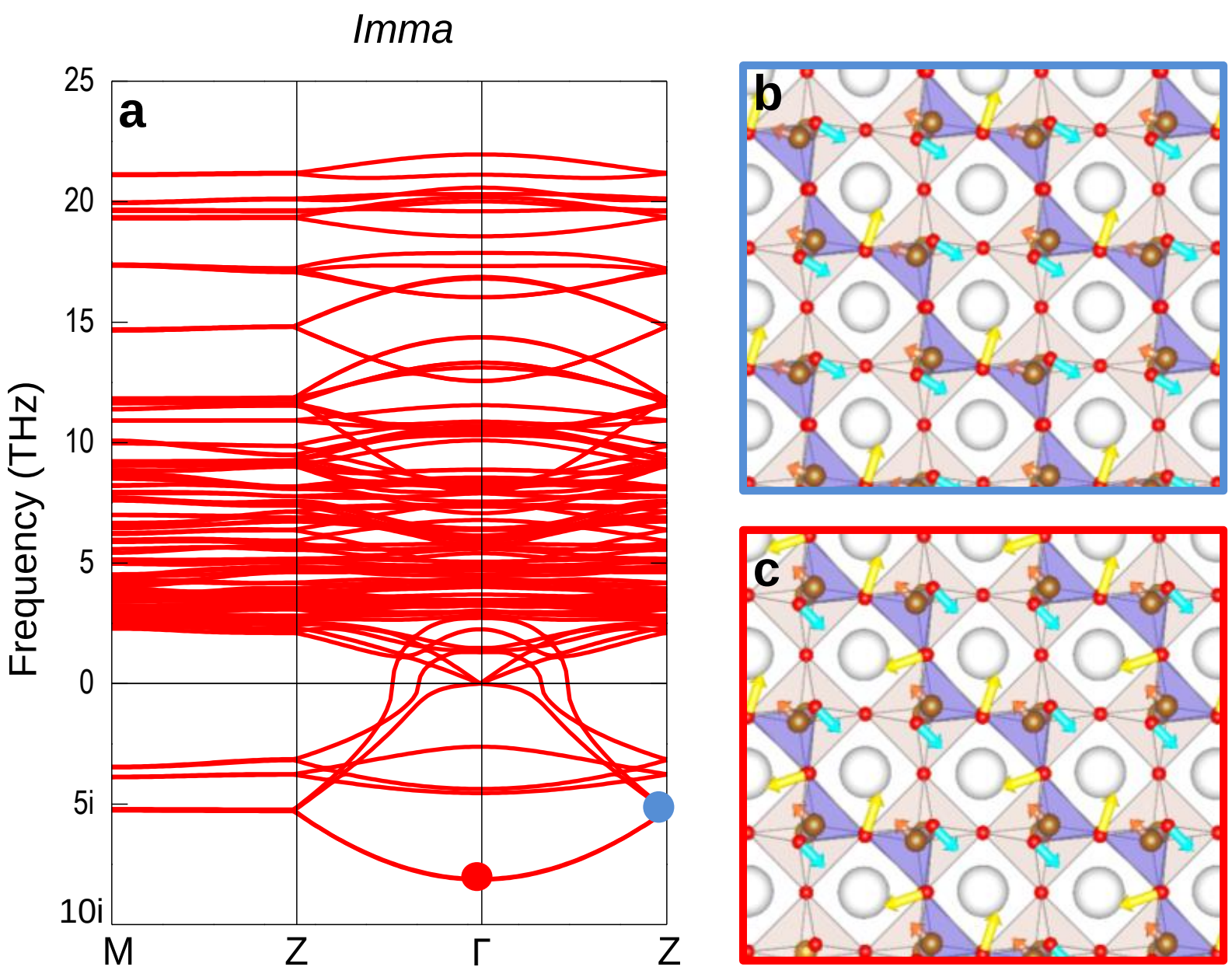

**Figure S3.** CPD examined by the phonon dispersion. (a) Phonon dispersion relations for an unstable *Imma* $SrFeO_{2.5}$ structure. Phonon modes for the lowest energy band dispersion at (b) the Z- and (c) Γ-points in the *Imma* $SrFeO_{2.5}$. The yellow and orange arrows represent the largest and smallest displacements, respectively.

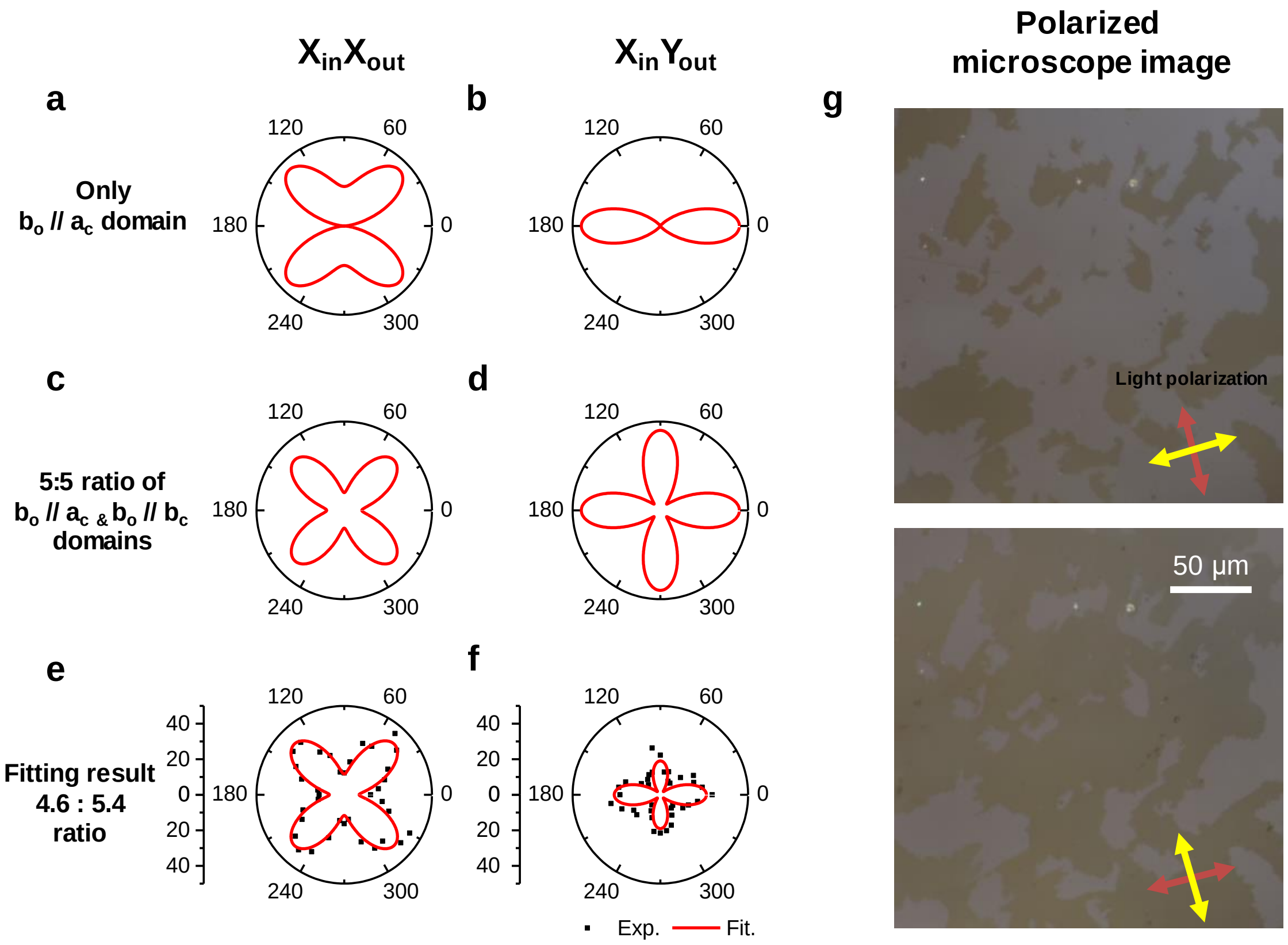

**Figure S4.** Domain structures of the $SrFeO_{2.5}$ epitaxial thin film. (a and b) SHG simulation results assuming that the $SrFeO_{2.5}$ thin film has $b_o$ // $a_c$ orientation for the $X_{in}$-$X_{out}$ and $X_{in}$-$Y_{out}$ configurations, respectively. (c and d) SHG simulation results assuming that the $SrFeO_{2.5}$ thin film has a one-half domain of $b_o$ // $a_c$ orientation and the other half-domain of $b_o$ // $b_c$ orientation for the $X_{in}$-$X_{out}$ and $X_{in}$-$Y_{out}$ configurations, respectively. (e and f) SHG simulation results assuming that the $SrFeO_{2.5}$ thin film has a 4.6:5.4 ratio of $b_o$ // $a_c$ and $b_o$ // $b_c$ orientations (red lines) for the $X_{in}$-$X_{out}$ and $X_{in}$-$Y_{out}$ configurations, respectively. These results match well with the experimental SHG results (dots). (g) The polarized optical microscopy images of the $SrFeO_{2.5}$ epitaxial thin film surface. By rotating the polarization direction by 90°, we can observe the reversed color contrast of the structural domains.

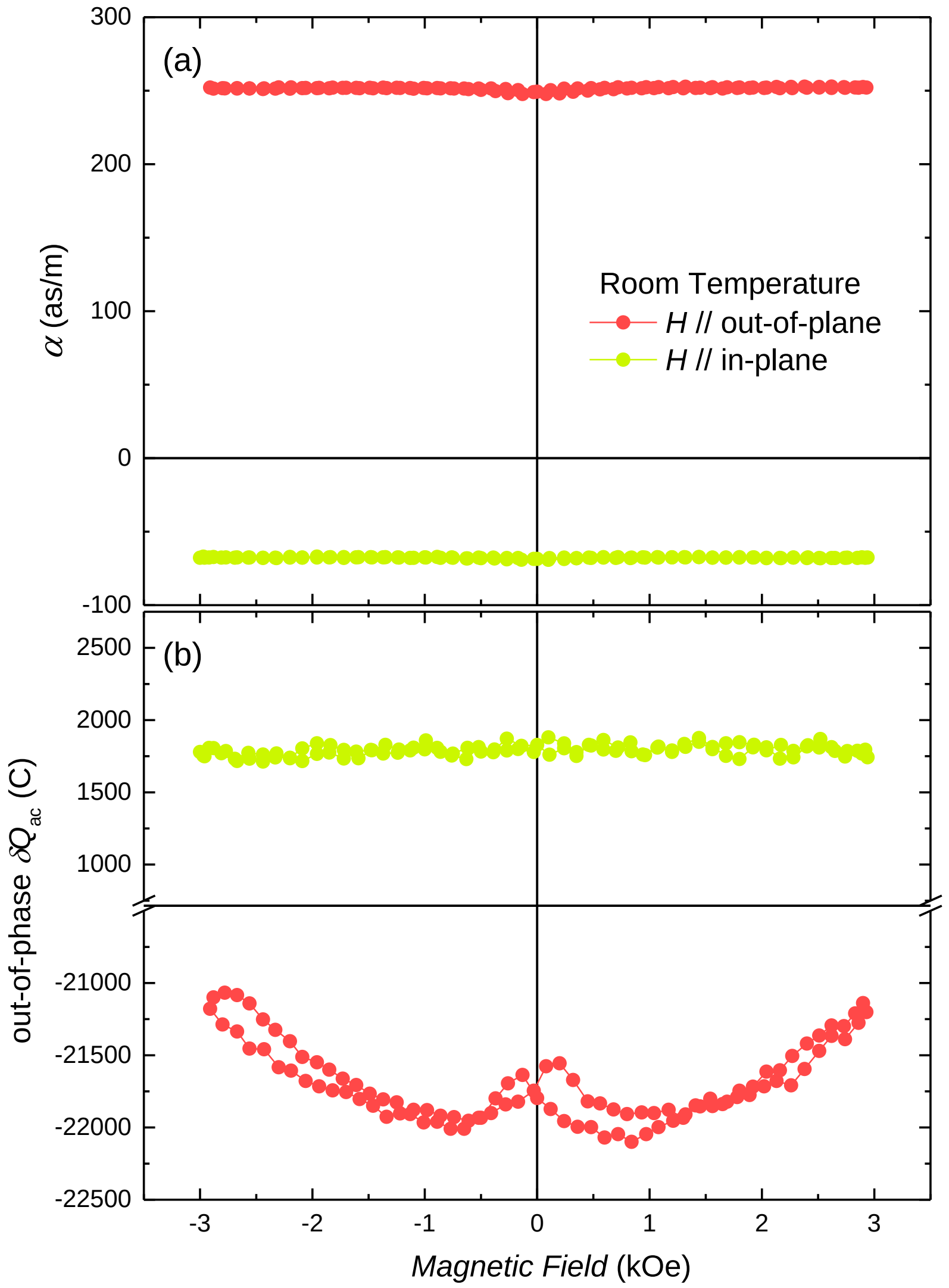

**Figure S5.** (a) DC magnetic field dependence of magnetoelectric coupling constant $\alpha(H) \equiv \delta P_{ac}/\delta H_{ac}$ at room temperature, when both DC and AC magnetic fields are applied along the out-of-plane (red) and in-plane (green) directions. (b) The corresponding out-of-phase signal of AC charge modulation $\delta Q_{ac}$, which is contributed from the eddy current by the oscillating magnetic field according to the Faraday's law. The magnetoelectric coupling constant $\alpha$ is estimated from the in-phase signal of $\delta Q_{ac}$ by ruling out the eddy current contribution to $\alpha$.